\documentclass[final]{aipproc}

\layoutstyle{8x11double}

\begin{document}

\title{The Saga of Landau-Gauge Propagators: \\[2mm] Gathering New Ammo}


\author{Attilio Cucchieri}{
  address={Instituto de F\'\i sica de S\~ao Carlos, Universidade de S\~ao Paulo \\[2mm]
           Caixa Postal 369, 13560-970 S\~ao Carlos, SP, Brazil}
}

\author{Tereza Mendes}{
  address={Instituto de F\'\i sica de S\~ao Carlos, Universidade de S\~ao Paulo \\[2mm]
           Caixa Postal 369, 13560-970 S\~ao Carlos, SP, Brazil}
}

\begin{abstract}
Compelling evidence has recently emerged from lattice
simulations in favor of the massive solution of the
Schwinger-Dyson equations of Landau-gauge QCD. The main objections to these
lattice results are based on possible Gribov-copy effects. We
recently installed at IFSC-USP a new GPU cluster dedicated to the study of Green's functions.
We present here our point of view on the Saga and the status of our project.
We also show data for the 2D case on a $2560^2$ lattice.
\end{abstract}

\maketitle

\section{The Saga}

Intuitively, the explanation of color confinement should be encoded
in the infrared (IR) behavior of QCD Green's functions. The
Landau-gauge Gribov-Zwanziger (GZ) confinement scenario and the scaling
solution obtained by solving Schwinger-Dyson equations (SDEs)
demand --- for any space-time dimensions D $\geq 2$ --- a null
gluon propagator at zero momentum and an IR-enhanced ghost
propagator \cite{Cucchieri:2008yp}.
At present, there is wide agreement \cite{Cucchieri:2010xr} that numerical
simulations in minimal Landau gauge show (in the infinite-volume limit):
{\bf 1)} an IR-finite gluon propagator $D(p)$ in D $=3,4$ and a null $D(0)$
in 2D, {\bf 2)} violation of reflection positivity for
the gluon propagator in D $=2,3,4$ and {\bf 3)} an essentially free
ghost propagator $G(p)$ in D $=3,4$ but IR-enhanced in 2D. Thus,
the 3D and 4D results support the massive solution of the SDEs
\cite{Boucaud:2008ji,massive}, while the 2D case has a scaling behavior.
Then, a natural question is why the 2D case is different from the 3D and 4D
ones. At the moment, a possible answer to this question has only been
presented in \cite{Dudal:2008xd}.

Recently, three works \cite{Sternbeck:2008mv,Maas:2008ri,Maas:2009se}
have allegedly shown evidence of the scaling IR behavior also in 3D and 4D.
Here, we will comment on these three works.

\subsection{The $\beta = 0$ Case}

We have already criticized Ref.\ \cite{Sternbeck:2008mv} in our
work \cite{Cucchieri:2009zt}. Since that criticism has not been answered,
we will repeat it here. The authors of Ref.\ \cite{Sternbeck:2008mv}
study the Landau-gauge gluon and ghost propagators in the strong-coupling limit of pure
SU(2) lattice gauge theory. These propagators are evaluated using
different discretizations of the gluon field and, in particular,
the standard (compact) definition and the (non-compact)
stereographic projection \cite{vonSmekal:2007ns}. Their main conclusions are:
``We furthermore demonstrate that the massive branch observed for $a^2q^2 <1$
does depend on the lattice definition of the gluon fields, and that it is thus
not unambiguously defined....One might still hope that this ambiguity
will go away at non-zero $\beta$ in the scaling limit. While this is true
at large momenta, we demonstrate...that the ambiguity is still present
in the low-momentum region, at least for commonly used values of the lattice
coupling such as $\beta = 2.3$ or $\beta = 2.5$ in $SU(2)$....The scaling
properties such as exponent and coupling, on the other hand, appear to be robust under
variations of the discretization of the gauge fields...This emphasizes the importance
of understanding any discretization ambiguity of the associated gluon mass, before concluding
that this mass is now firmly established.'' However, nowhere in Ref.\ \cite{Sternbeck:2008mv}
are data at $\beta = 2.5$ shown. On the
other hand, data for a lattice volume $32^4$ at $\beta = 2.5$ in the SU(2) case
are presented in Ref.\ \cite{vonSmekal:2007ns} for the two propagators, using the standard
discretization and the stereographic projection. The conclusion of \cite{vonSmekal:2007ns}
is that ``...there are hardly any differences between the propagators obtained in
each case''. Thus, referring to the last sentence reported above from Ref.\ \cite{Sternbeck:2008mv},
there are no discretization ambiguities in the evaluation of these propagators
and the existence of a gluon mass is now firmly established.

\subsection{The Absolute Landau Gauge}

Ref.\ \cite{Maas:2008ri} considers the absolute Landau gauge,
i.e.\ configurations belonging to the fundamental modular region
$\Lambda$. This approach, however, cannot yield an IR-enhanced ghost propagator
in 3D or in 4D. Actually, restricting the configuration space to the region $\Lambda$
makes the ghost propagator even less singular \cite{Cucchieri:1997dx}. This can be seen,
indeed, also in Figures 5 and 12 of Ref.\ \cite{Maas:2008ri}. The author tries
to explain these results, which clearly go against the scaling solution, by saying that
``The reason for this behavior of the ghost propagator...may be connected to
the volume evolution of the first Gribov region and the fundamental modular region....The
combined effect of the precise shape of the
low-eigenvalue spectrum and a diverging normalization of the eigenstates could be
sufficient to provide a more infrared divergent ghost propagator in the infinite-volume
and continuum limits in absolute Landau gauge than in minimal Landau gauge.''
Thus, simulations in the absolute Landau gauge should agree with the scaling solution in the
infinite-volume limit (for D $=3,4$) due to a {\em hypothetical} diverging contribution of
the eigenstates of the Faddeev-Popov operator. (Note that a possible way of quantifying
this sentence would be to prove that the lower bound of the ghost propagator, introduced
in Ref.\ \cite{Cucchieri:2008fc},
blows up sufficiently fast in the infinite-volume limit.) Moreover, that this effect should be important
in the absolute Landau gauge and not in the minimal Landau gauge remains a mystery to us,
considering that any configuration belonging to the absolute Landau gauge is
also a configuration of the minimal Landau gauge. The author also adds ``A final proof is, of course,
only that in the absolute Landau gauge at sufficiently large volume the ghost propagator would be more
singular than in the minimal Landau gauge. The volume dependence of the propagator in both gauges found
here is as expected if this is the case....Hence, at the current point it seems more appropriate
to compare the lattice results in absolute Landau gauge, rather than in minimal Landau gauge, with those
from functional results which exhibit a scaling behavior in the far infrared.'' These
statements --- which are somewhat {\em sibylline}, since the data do not show an
IR-enhanced ghost propagator ---
may be the reason why Ref.\ \cite{Maas:2008ri} is (wrongly) cited as a numerical
verification of the scaling behavior for D $ > 2$.

\begin{figure}
  \includegraphics[height=.3\textheight]{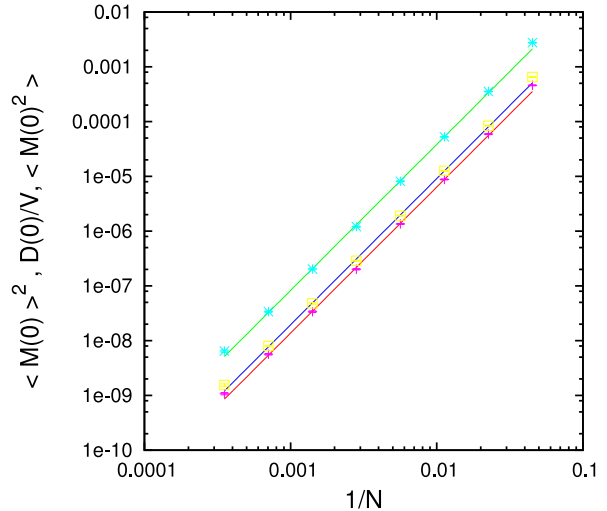}
  \caption{\label{fig:D0-2d}
           Plot of $D(0)/V$, together with its upper and lower
           bounds \cite{Cucchieri:2007rg}, as a function of the inverse
           lattice side $1/N$. In the three cases we get a behavior
           $1 / N^e$ with $e = 2.67(2)$.
           \vspace{-0.45mm}}
\end{figure}

\subsubsection{The 2D Case}

Let us note that if the massive behavior observed in 3D and in 4D could be
related to discretization effects, as suggested by Ref.\ \cite{Sternbeck:2008mv},
or to Gribov-copy effects, as reported in \cite{Maas:2008ri}, then these effects
should also be present for D $ = 2$ and one should not find in this case
a scaling behavior. In this respect, still in Ref.\ \cite{Maas:2008ri},
the author makes the following prediction: ``A consequence of this scenario is
that it should be expected that also in two dimensions, for sufficiently large
volumes and number of Gribov copies, an infrared finite gluon propagator is
obtained in the minimal Landau gauge.'' We checked this prediction by
evaluating the gluon propagator for lattice volumes up to $2560^2$ at $\beta = 10$
(i.e.\ with a lattice size $L \approx 460 \, fm$).
In Fig.\ \ref{fig:D0-2d} we plot the volume dependence of $D(0)/V$, together with the
upper and lower bounds introduced in Ref.\ \cite{Cucchieri:2007rg}. The three
sets of data clearly extrapolate to zero faster than $1/V$, implying $D(0)=0$ in 2D.

\subsection{The B Gauges}

After considering the absolute Landau gauge in Ref.\ \cite{Maas:2008ri}, a
new set of gauges --- called B gauges --- was introduced by the same author
\cite{Maas:2009se}.
In this case one looks along each orbit for a transverse configuration
that yields a given value $B$ for the ghost dressing function $D_G(p) = p^2 G(p)$ at
the smallest non-zero momentum $p_{min}$. This definition does not solve the Gribov
ambiguity \cite{Maas:2009se}. Moreover, in order to find an IR-enhanced ghost
propagator one needs to favor configurations closer to the
first Gribov horizon $\partial \Omega$. This is the opposite
of what is done in the absolute Landau gauge, where one favors configurations well
inside the first Gribov region $\Omega$ \cite{Cucchieri:1997ns}. Thus, if the B gauges
should produce the scaling solution, ``it could well be that...the absolute
Landau gauge is not connected to a scaling behavior'' \cite{Maas:2009se},
in disagreement with Ref.\ \cite{Maas:2008ri}. 

The main result of this approach is that the ghost propagator is strongly affected
by the choice of configuration on each orbit, in such a way that its values are enclosed in
a ``corridor''. In particular, in 3D and 4D the upper bound of this corridor
``is strongly increasing with volume''. At the same time, the gluon propagator seems
to be $B$-independent and we should have $D(0) > 0$ in the infinite-volume limit.
Thus, the only scaling solution that can be obtained with the B gauges seems to be the one
corresponding to a critical exponent $\kappa = 1/2$, which was never the
preferred value in scaling-solution works. Moreover, if the
infrared exponent is $1/2$ then in 4D one should have $ D_G(p) \sim 1/p$. Since
$1/p_{min} \approx L$, the upper bound of the corridor should grow at least as fast
as the lattice size $L$, in order to support the scaling solution. One can verify that
this is not the case with the 4D data presented in Fig.\ 3 of \cite{Maas:2009se}
(the curve should be hyperbolic as a function of $1/L$).

Let us note that one of the motivations for the introduction of the B gauges is the possible
relation with the one-parameter family of solutions obtained by functional methods
\cite{Boucaud:2008ji,Fischer:2008uz}. In this respect one should stress, however, that the B gauges are
related to different Gribov copies on each orbit. On the other hand, the configuration space
is not encoded in the SDEs and this information has to be put in by hand. This can be done
in simple cases \cite{Reinhardt:2008ij}, if all Gribov copies are known, but nobody
knows how to do it in a realistic case. Thus, this relation seems at the moment
quite accidental. Even more fanciful seems to us the hypothetical
connection between the Kugo-Ojima (KO) approach \cite{Cucchieri:2008yp} and a (possible) scaling
solution obtained using B gauges. Indeed, this connection requires ``subtle cancellation''
\cite{Maas:2009se}, since one has to relate an average over all
Gribov copies to results obtained by selecting specific copies inside the first Gribov region
$\Omega$. In our opinion, the lack of BRST invariance when the functional space is restricted
to $\Omega$ \cite{Dudal:2009xh} obscures the relation of the GZ approach 
with the KO criterion and the analogies between these two approaches seem to be, at
the moment, a questionable coincidence.

Finally, several questions should be answered before discussing
in detail the results obtained using B gauges. For example,
it is well known that some Gribov copies on the lattice are just lattice artifacts. Thus, by
using the B gauges, aren't we just probing these artifacts? This may explain the
over-scaling observed in \cite{Maas:2009ph}. Also, it seems very difficult to control the
infinite-volume limit of the corridor and, as pointed in \cite{Maas:2009se}, ``it cannot be
excluded that the corridor closes again at much larger volumes''.
This seems indeed possible since, for very large lattice volumes,
all the orbits should come very close to the boundary $\partial \Omega$ and one can
expect smaller Gribov-copy effects \cite{Sternbeck:2005tk}.

\section{Conclusions: the Ghost Factory}

We believe that, in order to understand the results obtained in minimal
Landau gauge using numerical simulations, the first question to be answered
is: why is the 2D case different? One could also ask: can we test
numerically the explanation presented in \cite{Dudal:2008xd}? A clear answer
to these questions probably requires new ideas and better data (especially
in the ghost sector). Unproven hypotheses and happy coincidences
should on the contrary be treated with great caution.

We recently installed at IFSC--USP a new machine with 18 CPUs Intel quadcore
Xeon 2.40GHz (with InfiniBand network and a total of 216 GB of memory) 
and 8 NVIDIA Tesla S1070
boards (500 Series), each with 960 cores and 16 GB of memory. The peak performance
of the 8 Tesla boards is estimated in about 2.8 Tflops in double precision and 33
Tflops in single precision. This machine will be used mainly for studies of Green's
functions in different gauges (Landau, Feynman and Coulomb) for various SU($N_c$) gauge groups.
In particular, the GPUs will be used for the inversion of the Faddeev-Popov matrix
using conjugate gradient.
This computer will allow us to perform an extensive study of the ghost sector.
We believe that this new ammo will help us clarify the issues addressed above.

\begin{theacknowledgments}
We thank FAPESP (grant \# 2009/52213-2) and CNPq for support.
\end{theacknowledgments}

\bibliographystyle{aipproc}

\end{document}